\documentstyle[prl,aps,epsfig]{revtex}         
\begin{document}
\catcode`\ä = \active \catcode`\ö = \active \catcode`\ü = \active
\catcode`\Ä = \active \catcode`\Ö = \active \catcode`\Ü = \active
\catcode`\ß = \active \catcode`\é = \active \catcode`\è = \active
\catcode`\ë = \active \catcode`\ô = \active \catcode`\ê = \active
\catcode`\ø = \active \catcode`\ò = \active \catcode`\í = \active
\defä{\"a} \defö{\"o} \defü{\"u} \defÄ{\"A} \defÖ{\"O} \defÜ{\"U} \defß{\ss} \defé{\'{e}}
\defè{\`{e}} \defë{\"{e}} \defô{\^{o}} \defê{\^{e}} \defø{\o} \defò{\`{o}} \defí{\'{i}}
\draft               
\twocolumn[\hsize\textwidth\columnwidth\hsize\csname
@twocolumnfalse\endcsname \vspace{-.4cm}
\title{Formation and Decay of Vortex Lattices in Bose-Einstein Condensates at Finite Temperatures}
\vspace{-.4cm}
\author{J.R. Abo-Shaeer$^{*}$, C. Raman$^{\dag}$, and W. Ketterle}
\address{Department of Physics, MIT-Harvard Center for Ultracold
Atoms, and Research Laboratory
of Electronics, \\
MIT, Cambridge, MA 02139}
\date{\today}
\maketitle

\begin{abstract}
The dynamics of vortex lattices in stirred Bose-Einstein
condensates have been studied at finite temperatures. The decay of
the vortex lattice was observed non-destructively by monitoring
the centrifugal distortions of the rotating condensate. The
formation of the vortex lattice could be deduced from the
increasing contrast of the vortex cores observed in ballistic
expansion. In contrast to the decay, the formation of the vortex
lattice is insensitive to temperature change.
\end{abstract}
 \vskip1pc ]

\vspace{-.4cm}

Gaseous Bose-Einstein condensates (BEC) have become a testbed for
many-body theory. The properties of a condensate at zero
temperature are accurately described by a nonlinear
Schr\"{o}dinger equation. More recently, theoretical work on the
ground state properties of condensates \cite{dalf99rmp} has been
extended to rotating condensates containing one or several
vortices \cite{fett01} and their dynamics.  These include vortex
nucleation \cite{dalf01crit,angl01surf}, crystallization of the
vortex lattice \cite{tsub01}, and decay \cite{zhur00,fedi00}.
Experimental study has focused mainly on the nucleation of
vortices, either by stirring condensates directly with a rotating
anisotropy \cite{madi00,abos01latt,hodb01vort} or creating
condensates out of a rotating thermal cloud \cite{halj01}. Here we
report the first quantitative investigation of vortex dynamics at
finite temperature.  The crystallization and decay of a vortex
lattice have been studied and a striking difference is found
between the two processes:  while the crystallization is
essentially temperature independent, the decay rate increases
dramatically with temperature.

The method used to generate vortices has been outlined in previous
work \cite{abos01latt,rama01nuc}. Condensates of up to 75 million
sodium atoms ($> 80 \%$ condensate fraction) were prepared in a
cigar-shaped Ioffe-Pritchard magnetic trap using evaporative
cooling. The radial and axial trap frequencies of $\omega_x =
2\pi~\times~(88.8 \pm 1.4)$ Hz, $\omega_y = 2\pi~\times~(83.3 \pm
0.8)$ Hz, and $\omega_z = 2\pi~\times~(21.1 \pm 0.5)$ Hz,
respectively, corresponded to a radial trap asymmetry of
$\epsilon_r = (\omega_x^2-\omega_y^2)/(\omega_x^2+\omega_y^2) =
(6.4 \pm 0.2) \%$ and an aspect ratio of $A=\omega_x/\omega_z =
(4.20 \pm 0.04)$. The relatively large value of the radial trap
asymmetry is due to gravitational sag and the use of highly
elongated pinch coils (both estimated to contribute equally). The
radio frequency used for evaporation was held at its final value
to keep the temperature of the condensate roughly constant
throughout the experiment. The condensate's Thomas-Fermi radii,
chemical potential, and peak density were $R_{r} = 28~\mu$m,
$R_{z} = 115~\mu$m, 300 nK, and $4 \times 10^{14}$ cm$^{-3}$,
respectively, corresponding to a healing length $\xi \simeq 0.2$
$\mu{\rm m}$.

Vortices were produced by spinning the condensate for 200 ms along
its long axis with a scanning, blue-detuned laser beam (532 nm)
\cite{madi00,onof00}. For this experiment two symmetric stirring
beams were used (Gaussian waist $w$ = 5.3 $\mu$m, stirring radius
24 $\mu$m). The laser power of 0.16 mW per beam corresponded to an
optical dipole potential of 310 nK. After the stirring beams were
switched off, the rotating condensate was left to equilibrate in
the static magnetic trap for various hold times. As in our
previous work, the vortex cores were observed using optical
pumping to the $F=2$ state and resonant absorption imaging on the
$F$ = 2 to $F$ = 3 cycling transition \cite{abos01latt}. After
42.5 ms of ballistic expansion the cores were magnified to 20
times their size, $\xi$, in the trap.

The decay of the vortex lattice can be observed by allowing the
condensate to spin down in the trap for variable times before the
ballistic expansion and subsequent observation of the vortex
cores. However, this method requires destructive imaging and
suffers from shot-to-shot variations. \textit{In situ}
phase-contrast imaging is non-destructive, but the vortex cores
are too small to be resolved.  Instead, we monitored the
centrifugal distortion of the cloud due to the presence of
vorticity. This is a quantitative measure for the rotation
frequency, $\Omega$, of the lattice, and therefore the number of
vortices~\cite{nozi90}. Such distorted shapes have been observed
previously for rotating clouds\cite{abos01latt,halj01,rama01nuc}.

The shape of a rotating condensate is determined by the  magnetic
trapping potential and the centrifugal potential
$-\frac{1}{2}M\Omega^2r^2$, where $M$ is the atomic mass.  From
the effective radial trapping frequency $\omega_r \rightarrow
\omega_r' = \sqrt{\omega_r^2-\Omega^2}$, one obtains the aspect
ratio of the rotating cloud, $A'$ = $\omega_r'/\omega_z$, as

\begin{equation}
A' = A\sqrt{1-(\Omega/\omega_r)^2}.\label{eq:AspectRatio}
\end{equation}

\begin{figure}[htbf]
\begin{center}
\epsfxsize=85mm {\epsfbox{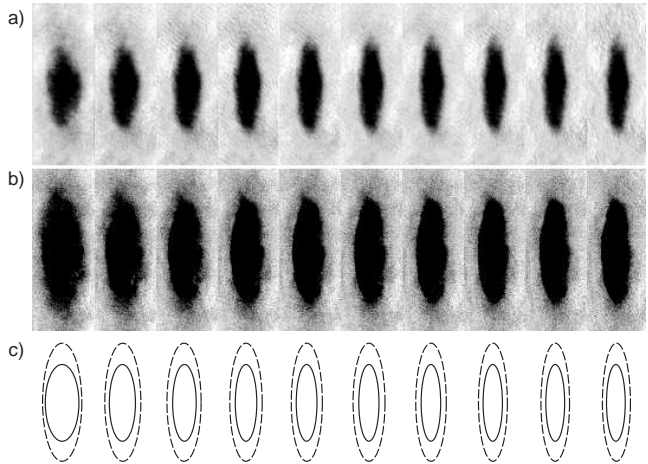}}
\end{center}
\caption{a) Spin down of a rotating condensate in a static
magnetic trap. The first phase contrast image is taken 100 ms
after turning off the drive, with each subsequent image spaced by
100 ms. The rotating vortex lattice caused a radial swelling of
the condensate, reducing the aspect ratio. As vortices leave the
system the aspect ratio approaches its static value of 4.20. The
field of view is 25 $\mu$m by 75 $\mu$m. b) Observation of a
rotating thermal cloud.  The parameters are identical to (a), but
the probe light detuning is closer to resonance, enhancing the
sensitivity to the more dilute thermal cloud. The phase shift of
the dense condensate exceeds 2$\pi$ and is displayed as saturated
in the image. The apparent loss in atom number is due to Rayleigh
scattering of the probe light. c) The inner and outer contours
represent the aspect ratio of the condensate and thermal cloud,
respectively, as obtained from two-dimensional fits to
phase-contrast images.} \label{fig:DecayStrip}
\end{figure}

For the decay measurements, the condensate was stirred for 200 ms,
producing $\sim$ 130 vortices on average at the coldest
temperature (determined using time-of-flight imaging). After the
drive was stopped the cloud equilibrated in the stationary
magnetic trap. Ten \textit{in situ} images of each condensate were
taken at equal time intervals using non-destructive phase contrast
imaging detuned 1.7~GHz from resonance (see
Fig.~\ref{fig:DecayStrip}a). We verified that the decay was not
affected due repeated imaging by varying the time between
exposures and observing no difference. The theoretical treatment
of the decay has two limiting cases, one where the thermal cloud
is nearly stationary, the other where the thermal cloud is closely
following the rotating condensate~\cite{zhur00}. Because the
thermal cloud could not easily be discerned in images taken with
1.7~GHz detuning, we used light closer to resonance (400 MHz
detuning) to determine the shape of the thermal cloud
(Fig.~\ref{fig:DecayStrip}b).

\vspace{-.1cm}

The aspect ratio of the condensate was obtained by fitting a 2D
Thomas-Fermi profile to the 1.7 GHz images. The aspect ratio of
the thermal cloud was determined from the 400 MHz images by
masking off the condensate in the inner region of the cloud, and
then fitting a 2D Gaussian to the image.  The comparison of the
contours of the condensate and thermal cloud (condensate fraction
of $N_c/N$ = 0.62) shows that the thermal cloud is also rotating.
The images for the thermal cloud show that the aspect ratio decays
from 3.0 to 4.2 on the same time scale as the condensate.  This
corresponds to an initial rotation rate of $\sim$ 2/3 that of the
condensate.

The damping rate of the vortex lattice, obtained from exponential
fits to the data, was studied at different temperatures by varying
the condensed fraction of atoms in the trap.  The condensate
fractions were obtained from fits to time-of-flight images of the
condensate \textit{before} the stirring.  The temperature was
derived from these values using the scaling theory for an
interacting Bose gas~\cite{gior97jltp}.  At a fixed rf frequency
for evaporation, the centrifugal potential lowers the trap depth
by a factor $1-(\Omega/\omega_r)^2$ for a thermal cloud rotating
at frequency $\Omega$.  Evaporation should lower the temperature
of the rotating cloud by the same factor.

\begin{figure}[htbf]
\begin{center}
\epsfxsize=85mm{\epsfbox{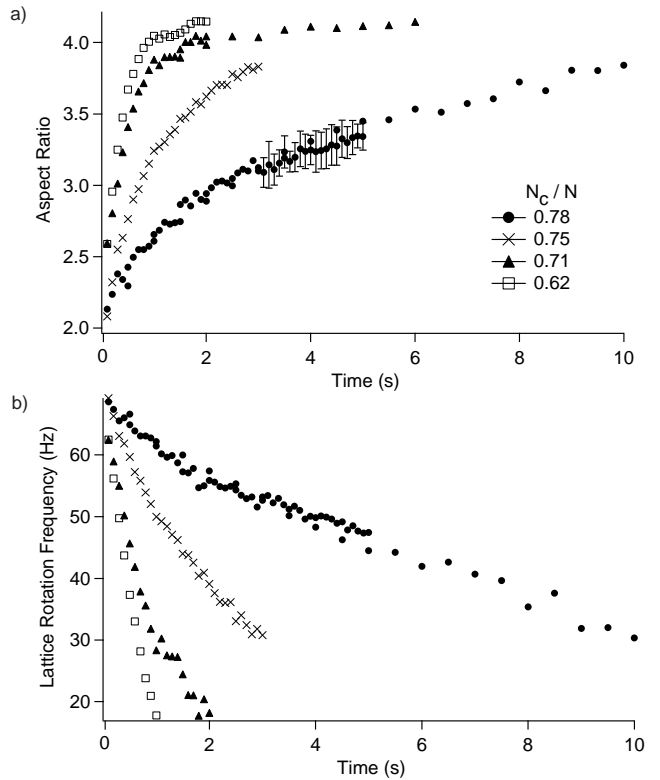}}
\end{center}
\caption{Decay of a vortex lattice at finite temperatures. a) The
aspect ratio of the condensate approaching its static value. Each
point represents the average of 10 measurements. Error bars given
for $(\bullet)$ are statistical and are typical of all four data
sets. b) The decay of the rotation rate of the lattice for the
same data using Eq.~(\ref{eq:AspectRatio}). The decay of the
rotation depends strongly on the thermal
component.}\label{fig:DecayGraphs}
\end{figure}

\begin{figure}[htbf]
\begin{center}
\epsfxsize=85mm{\epsfbox{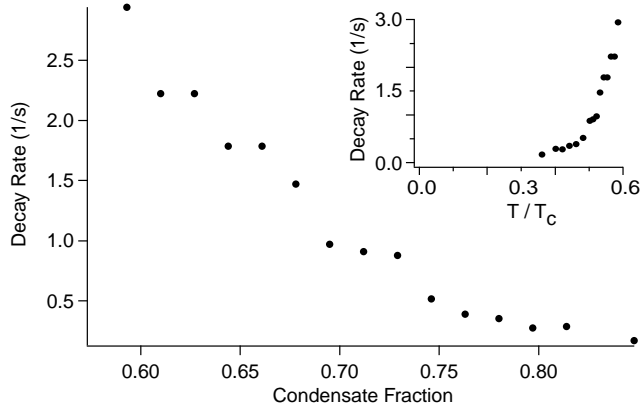}}
\end{center}
\caption{Decay rates for vortex lattices at several temperatures.
The rates are determined from data taken during the 1st second of
equilibration. Each point represents the average of 5
measurements. The inset figure emphasizes the strong dependence of
the decay rate on temperature, T. The transition temperature was
$T_{c}$ = 790 nK and varied by less than 7~\% for the data shown.}
\label{fig:DecayRate}
\end{figure}

Fig.~\ref{fig:DecayGraphs}a shows how the aspect ratio of the
condensate approaches its static value as the vortex lattice
decays.  The decrease of the angular speed appears to be
exponential (Fig.~\ref{fig:DecayGraphs}b) and strongly depends on
temperature (Fig.~\ref{fig:DecayRate}).

\vspace{-.1cm}

\begin{figure}[htbf]
\begin{center}
\epsfxsize=85mm {\epsfbox{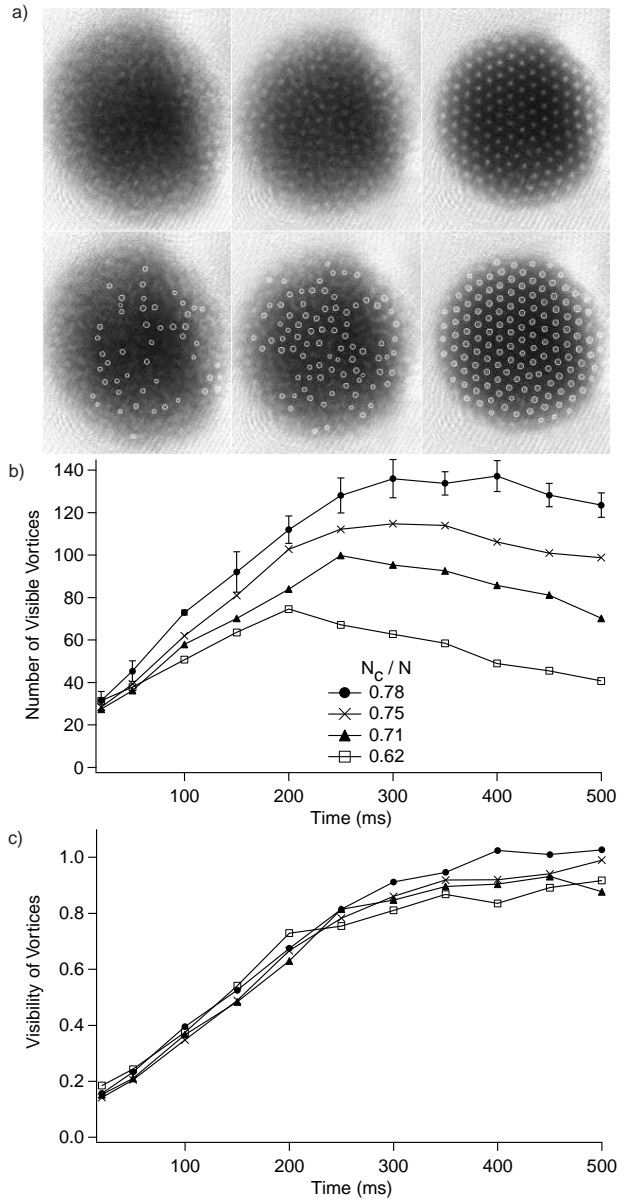}}
\end{center}
\caption{Crystallization of the vortex lattice.  a) The top row
shows three condensates that have equilibrated for 50, 150 and 300
ms, respectively, and have 48, 86 and 140 vortices recognized as
``visible'' by our algorithm. The bottom row shows the same
condensates with the ``visible'' vortices circled. The field of
view was 1.4 mm by 1.6 mm. b) Growth of the number of visible
vortices for several temperatures expressed by the condensate
fraction ($N_c/N$). c) Visibility of vortices derived from the
data in figure b, normalized by the number of vortices inferred
from centrifugal distortion measurements.} \label{fig:Formation}
\end{figure}

In addition to the decay process, the formation of the vortex
lattice has also been examined. After rotating the condensate it
typically took hundreds of ms for the lattice to form (see Fig. 4
in ref. \cite{abos01latt}). One may expect the lattice to already
form in the rotating frame during the stirring because the lattice
is the lowest energy state for a given angular momentum. This
absence of equilibration in the rotating frame is presumably due
to heating and excitation of collective lattice modes by the
stirring beams. The crystallization of the vortex lattice was
studied by determining the contrast or visibility of the vortex
cores as a function of equilibration time. To avoid any bias, we
used an automated vortex recognition algorithm. First, each image
was normalized by dividing it by a blurred duplicate of itself.  A
binary image was then created with a threshold set to the value at
which the algorithm reliably detected almost all vortices that
were identified by visual inspection of equilibrated images (there
was less than 5$\%$ discrepancy). Clusters of contiguous bright
pixels within a circular area were counted as ``visible''
vortices. Fig.~\ref{fig:Formation}a shows three vortex lattices
after different equilibration times with the visible vortices
identified.

\vspace{-.1cm}

The lowest temperature (337 nK, corresponding to a condensate
fraction of $\approx$ 80\%) was close to the chemical potential.
Further cooling would have resulted in smaller condensates.  The
highest temperature (442 nK) was limited by the rapid decay of the
vortex lattice.

Fig.~\ref{fig:Formation}b,c show that the formation of the lattice
depends very weakly on temperature, if at all, over the range
studied. The larger number of vortices crystallized at lower
temperatures (Fig.~\ref{fig:Formation}b) is due to the strong
temperature dependence of the lattice decay rates, which differed
by more than a factor of 5.  We can correct for this by estimating
the number of vortices $N_v(t)$ as a function of time from the
centrifugal distortion measurements described above as $N_v(t)=2
\kappa \Omega(t) \pi R(t)^2$, where $\kappa=h/M$ is the quantum of
circulation, $M$ the sodium mass and $R$ the radial Thomas-Fermi
radius.  By normalizing the number of visible vortices by this
estimate, we deduce the vortex visibility as a function of time.
These lattice formation curves overlap almost perfectly for
different temperatures (Fig.~\ref{fig:Formation}c).

\vspace{-.1cm}

The decay of a vortex lattice is discussed in
refs.~\cite{zhur00,fedi00}.  The process is modelled as a two-step
transfer of angular momentum. The rotation of the condensate is
damped due to friction with the thermal cloud ($\mu_{c-th}$),
which is in turn damped via friction with the trap anisotropy
($\mu_{th-\epsilon_{r}}$). Even at our highest temperatures,
$\mu_{c-th}$ is sufficiently larger than $\mu_{th-\epsilon_{r}}$
to spin up the thermal cloud to $\sim$ 2/3 of the lattice rotation
rate. $\mu_{th-\epsilon_{r}}$ is calculated in ref.~\cite{guer00}
for a classical Boltzmann gas. For our parameters the relaxation
time $\tau_{\rm rot}$ for the thermal cloud rotation is $\tau_{\rm
rot}=4 \tau$, where $\tau$ is the relaxation time for quadrupolar
deformations. $\tau$ is approximately $5/4$ of the time between
elastic collisions $\tau_{\rm el}= 2 / n v_{\rm th}\sigma$
\cite{guer99osc}, where $n$ is the atomic density, $v_{\rm th}$
the thermal velocity and $\sigma$ the elastic cross section. For a
quantum-saturated thermal cloud, this gives a relaxation time
proportional to $T^{-2}$.

The presence of the condensate adds additional moment of inertia,
extending the rotational relaxation time by a factor $f_{\rm
inertia}=(I_{\rm th}+I_c) /I_{\rm th}$, where $I_{\rm th}$ and
$I_c$ are the moments of inertia of the thermal cloud and the
condensate, respectively. At high temperatures this factor
approaches one, because the moment of inertia of the thermal cloud
is dominant. For very low temperatures $f_{\rm inertia} \sim
(I_{c}$ / $I_{th})$.  This ratio is \cite{zhur00}
\begin{equation}
\frac{I_c}{I_{\rm th}} = \frac{15 \zeta(3)}{16 \zeta(4)}
\frac{N_c}{N_{\rm th}} \frac{\mu}{k_B T} \label{eq:inertia}
\end{equation}
where the numerical pre-factor is 1.04.  Thus, for a
noninteracting Bose gas in the low temperature limit $f_{\rm
inertia}$ scales as $T^{-4}$ and the relaxation time of the vortex
lattice should scale as $T^{-6}$. The observed temperature
dependence of the relaxation time, decreasing by a factor of 17
for a 60\% increase in the temperature, agrees fortuitously well
given the approximations made in the theory. The absolute
relaxation rates predicted by refs.~\cite{zhur00,guer00} are on
the order of 1 $s^{-1}$, in reasonable agreement with our results.
However, for a quantitative comparison to theory it will be
necessary to further characterize the rotating thermal cloud,
which may not be in full equilibrium.

In contrast to the decay, the crystallization process of the
vortex lattice was essentially independent of temperature.  This
was unexpected because all dissipative processes observed thus far
in BECs, including the decay of the vortex lattice, have shown a
strong temperature dependence \cite{jin97,stam98coll}. D.~Feder
has numerically simulated a stirred condensate using only the
Gross-Pitaevskii equation. The results show that the absence of
dissipation leads to rapid, irregular motion of the vortices.  The
addition of a phenomenological dissipative term in
ref.~\cite{tsub01} resulted in the formation of triangular vortex
lattices.  However, the origin of the dissipation was not
identified. Our results suggest that the time-limiting step for
the evolution of a vortex tangle into a regular lattice does not
strongly depend on temperature. One possibility is that the
thermal cloud is not directly involved. This may be similar to the
reconnection of vortices or the damping of Kelvin modes, where the
spontaneous emission of phonons act as a dissipative mechanism
\cite{lead01,kivo01}. Another possibility is that the
rearrangement of the vortices into a rectilinear lattice is slow
and not limited by dissipation.

In conclusion, we have studied the crystallization and decay of
vortex lattices.  Both processes are dissipative and require
physics beyond the Gross-Pitaevskii equation.  The dynamics of
vortices nicely illustrates the interplay of experiment and theory
in the field of BEC.

The authors would like to acknowledge T. Rosenband for development
of the vortex recognition algorithm.  We also thank J.M. Vogels
and K. Xu for their experimental assistance and J.R. Anglin, Z.
Hadzibabic, A.E. Leanhardt, and R. Onofrio for insightful
discussions. This research is supported by NSF, ONR, ARO, NASA,
and the David and Lucile Packard Foundation.

\end{document}